\input harvmac
\newcount\figno
\figno=0
\def\fig#1#2#3{
\par\begingroup\parindent=0pt\leftskip=1cm\rightskip=1cm\parindent=0pt
\global\advance\figno by 1
\midinsert
\epsfxsize=#3
\centerline{\epsfbox{#2}}
\vskip 12pt
{\bf Fig. \the\figno:} #1\par
\endinsert\endgroup\par
}
\def\figlabel#1{\xdef#1{\the\figno}}
\def\encadremath#1{\vbox{\hrule\hbox{\vrule\kern8pt\vbox{\kern8pt
\hbox{$\displaystyle #1$}\kern8pt}
\kern8pt\vrule}\hrule}}
\def\underarrow#1{\vbox{\ialign{##\crcr$\hfil\displaystyle
 {#1}\hfil$\crcr\noalign{\kern1pt\nointerlineskip}$\longrightarrow$\crcr}}}
%
\overfullrule=0pt

%

\def\bar{\overline}

\def\R{{\bf R}}

\Title{ } {\vbox{\centerline{Quest For Unification}
\bigskip
}}
\centerline{Edward Witten\foot{Supported in part by NSF Grant
PHY-0070928.}}
\smallskip
\centerline{\it Institute For Advanced Study, Princeton NJ 08540 USA}


\medskip

\noindent
\vskip 2cm

Let us begin by recalling how the standard model of particle
physics looked before grand unification was proposed.\foot{This
article is based on my Heinrich Hertz lecture at SUSY 2002 at
DESY, June, 2002.} Focus, in particular, on the fermions. The
left-handed fermions of one generation can be displayed as
follows: \eqn\onegen{\left(\matrix{\nu \cr e^-\cr}\right)_{-1},~
\left(\matrix{u \cr d\cr}\right)_{1/3},~\bar u_{-4/3},~\bar
d_{2/3},~e^+_2.} Here I have explicitly displayed the pairs
$(\nu,e^-)$ and $(u,d)$ as weak doublets, but I have not
explicitly shown the color quantum numbers of $u,d$ and $\bar
u,\bar d$. The subscript is the hypercharge quantum number $Y$.

This structure looks strange at first sight.  Part of the
strangeness, relative to the way physics was formerly understood,
is that the left- and right-handed fermions transform differently
under the standard model gauge symmetries.  With time, physicists
came to see this as a virtue: it means that the  quarks and
leptons cannot have bare masses and can only gain mass from the
Higgs mechanism. So it explains why they cannot be much heavier
than the $W$ and $Z$ bosons (which gain mass from the same Higgs
mechanism) and in particular why they are
 light compared to hypothetical fundamental scales in physics such as the Planck
scale of gravity.

But there is another side to the strangeness in the structure of
the standard model. A single generation of quarks and leptons is
made by putting together quite a few odd bits and pieces, with
strange fractions, in particular, for the hypercharge quantum
numbers.  Anomaly cancellation depends on obscure calculations
like
\eqn\ancan{\Tr\,Y^3=2(-1)^3+6(1/3)^3+3(-4/3)^3+3(2/3)^3+2^3=0.}
There is no virtue in this kind of strangeness.

So it was an illumination in 1973 when Georgi and Glashow
(building on prior work of Pati and Salam on quark-lepton
unification) unveiled the  $SU(5)$ grand unified theory, or GUT.
In this theory, all known elementary particle gauge forces were
interpreted as part of a single underlying $SU(5)$ gauge force.
  Indeed,
  $SU(5)$ is  the smallest and most obvious
simple or unified gauge group in which one can embed the standard
model gauge interactions. The embedding is made via a block
diagonal ansatz that is based on the fact that $3+2=5$:
\eqn\knoin{\left(\matrix{ SU(3) &
* \cr
                            *    &    SU(2)\cr}\right).}
(Hypercharge is generated in $SU(5)$ by
 a traceless diagonal matrix that
commutes with $SU(3)\times SU(2)$.) In this framework, a standard
model generation is reduced to two pieces, the ${\bf 10}$ and the
$\bf{\bar 5}$, schematically
\eqn\noin{\left(\matrix{0 & \bar u &
\bar u & u & d\cr
                        & 0 & \bar u & u & d\cr
                          &  & 0 & u & d\cr
                          &  &  &  0 & e^+\cr
                           &&&& 0 \cr}\right)~\bigoplus ~\left(\matrix{\bar d\cr
                              \bar d\cr \bar d\cr \nu \cr e^-\cr}\right).}
The fractions are explained as consequences of $SU(5)$ group
theory, and the verification of anomaly cancellation is greatly
abridged.   The standard model has never looked the same  since
the unified $SU(5)$ model was proposed.

To achieve these attractive results, it
was necessary, as we see in eqn. \noin, to place quarks and
leptons (and their antiparticles) in the same representations.  As in the model of
 Pati and Salam, this led to a prediction of
{\it proton decay}. In fact, the model predicts new gauge bosons
$X$ and $Y$, corresponding to the off-diagonal blocks in eqn.
\knoin; they mediate processes such as $p\to e^+\pi^0$.

There are a few immediate problems.   If the strong interactions
are unified with the weak and electromagnetic interactions, why
are they so much stronger?  And will the proton be sufficiently
long-lived?  These issues were addressed in 1974 in a celebrated
paper by Georgi, Quinn, and Weinberg, who calculated the
``renormalization group running'' of the strong, weak, and
electromagnetic couplings.  They found that unification was
possible, provided that the weak mixing angle had the right value
$\sin^2\theta_W\approx .20$ and the unification scale was about
$10^{15}$ GeV.  These were spectacular results, since the value
of the weak angle was about right, and the value of the
unification scale was fortuitous.  In fact, $10^{15}$ GeV was
close enough to the Planck scale to suggest a unification with
gravity, and big enough to make the proton long-lived.  A proton
lifetime of about $10^{30}$ years was predicted, large enough to
be compatible with experiment but small enough to be observable.
The model would have failed if the computed unification scale were
significantly less than about $10^{15}$ GeV (because the inferred
proton lifetime would have been too short) and would have been
implausible if the inferred unification scale were much greater
than $10^{19}$ GeV.

A GUT scale of $10^{15}$ GeV may not seem like it is really very
close to a Planck scale of $10^{19}$ GeV.  But this calculation
should really be viewed on a log scale, since renormalization
group running is logarithmic and what is computed is really the
logarithm of the GUT scale.  On a log scale, this early
computation did give something pretty close to the logarithm of
the Planck mass, and it was the first time that any sort of
particle physics computation gave a result for any characteristic
scale of particle physics phenomena that was at all close to the
Planck scale.

It was nice in $SU(5)$ to reduce the mess of a standard model generation to just two pieces.
But can one do better?  It was soon seen (by Georgi and by Fritzsch and Minkowski) that in
 the larger group $SO(10)$, all quarks and leptons                              of one generation
 fit neatly in a single irreducible representation.  There is a price, though: one has to add
 a left-handed anti-neutrino.  Since it is a standard model singlet, it is natural for it to get
 a GUT scale mass, and this led to the idea of the ``see saw'' mechanism for neutrinos
 (due to Gell-Mann, Ramond, and Slansky and to Yanagida). A model
 like this will, in the basis $\left(
 \matrix{\nu_L\cr \bar\nu_L\cr}\right)$, give
 a mass matrix for the left-handed neutrino and anti-neutrino of the form
 \eqn\onon{\left(\matrix{0 & m\cr m & M\cr}\right).}
  Here $m$ comes from the electroweak Higgs
 effect, and $M$ can be of order the unification scale $M_{GUT}$.  If we take $m$ to be of
 order $M_Z$ and $M=10^{15}$ GeV,
 we get a light neutrino mass in the range $m_\nu\sim M_Z^2/M\sim (10^2 \, {\rm GeV})/
 10^{15}\,{\rm GeV}=.01$
 eV.  This estimate, which was made in the late 1970's, gave a big impetus to the search for
 neutrino masses and oscillations.  Of course, in making this estimate, it is not clear just
 what we should put in either the numerator or the denominator, and since experiments of twenty
 years ago were not sensitive to neutrino masses of $.01$ eV, there was considerable interest
 in variants in which the neutrino masses would be somewhat bigger, for example because $M$ arises
 from loop effects and is smaller than $M_{GUT}$.

 Having come this far, can we go farther and (i) unify {\it three} generations of quarks and
 leptons in {\it one} irreducible representation of a larger gauge group? or (ii) unify
 Higgs bosons with gauge fields, or with quarks and leptons?  The answer to these questions
 was ``no'' in the framework of four dimensional grand unification.

 For example, there is no
 candidate grand unified or GUT gauge group that puts several {\it chiral} families in one
 irreducible representation -- without ``antifamilies'' of the opposite chirality.  Of course,
 experimental bounds on such antifamilies have become progressively tighter, and from a theoretical
 point of view, the possibility that they would combine with the ordinary families and get a large
 bare mass makes them seem unattractive.

Likewise, in four-dimensional GUT's, one cannot really unify Higgs particles with quarks and
leptons or gauge bosons (or for that matter, unify quarks and leptons with gauge bosons).
The closest try uses supersymmetry plus the $E_6$ model that I will get to shortly.

The other recognized problem of GUT's in this period was the
``gauge hierarchy problem,'' which is the question of why
$SU(2)\times U(1)$ breaking is so weak compared to GUT symmetry
breaking.  In the context of the $SU(5)$ model, why are $M_W$ and
$M_Z$ so much less than $M_X$ and $M_Y$?

In this brief review of the GUT theories of the 1970's, I have stressed the $SU(5)$ and $SO(10)$
models, which made sense of the fermion quantum numbers and led to predictions of
proton decay and neutrino masses.  Are there any bigger groups that can teach us more?

Personally, I would say that there is no four-dimensional GUT
model that does better -- but there is one more model worthy of
note.  This is the $E_6$ model, introduced by Gursey, Ramond, and
Sikivie.  What is $E_6$?  In gauge theory, we need to pick a
gauge group, which, if we wish to achieve unification, should be
a simple gauge group, and moreover should be compact (so that all
gauge bosons have positive kinetic energy).  There is a nice
classification of these groups.  First, there are three infinite
families, $SU(N)$, $SO(N)$, and $Sp(N)$.  Two groups from these
infinite families -- $SU(5)$ and $SO(10)$ -- are used in the
models that I have mentioned so far. Describing nature by a group
taken from an infinite family does raise an obvious question --
why this group and not another?  In addition to the three infinite
families, there are five exceptional Lie groups, namely
$G_2,F_4,E_6,E_7$, and $E_8$. Since nature is so exceptional, why
not describe it using an exceptional Lie group?

Of the five exceptional Lie groups, four  ($G_2$, $F_4$, $E_7$,
and $E_8$) only have real or pseudoreal representations. A
four-dimensional GUT model based on such a group will not give
the observed chiral structure of weak interactions. The one
exceptional group that does have complex or chiral representations
is $E_6$, and this one works beautifully. The grand unified theory
based on $E_6$ is not clearly superior to the $SO(10)$ model, but
it does capture the successes of the $SO(10)$ model
``exceptionally.''

But why $E_6$?  The exceptional groups fit into a chain of embeddings
\eqn\onon{G_2\subset F_4\subset E_6\subset E_7\subset E_8.}
If nature likes exceptional groups, why stop half-way?  Yet $E_6$ is the only
exceptional group that works for four-dimensional GUT's.

There is another interesting chain of group embeddings
(popularized notably  by Olive), \eqn\bonon{SU(5)\subset
SO(10)\subset E_6\subset E_7\subset E_8.} Here, at each step, one
adds another node at one end of the Dynkin diagram. It is notable
that the most significant GUT models correspond to the first three
gauge groups in this chain.  But again, we have the same question
that we asked in connection with eqn. \onon: if nature likes this
chain, why stop half-way?  As before, in four-dimensional GUT's,
we can only go half-way down this chain because of the $V-A$
structure of weak interactions.

The next developments that I will mention were experimental.  More accurate measurements
showed that $\sin^2\theta_W$ is close to the GUT value, but not close enough.
And the proton lifetime turned out to be longer than predicted in the simplest GUT's.

Both of these problems were neatly addressed (by Dimopoulos, Raby, and Wilczek) by repeating
the Georgi-Quinn-Weinberg calculation in the presence of supersymmetry.  Supersymmetry raises
the GUT prediction for $\sin^2\theta_W$, which becomes very close to the modern measurement.
It also raises the GUT scale, making the proton lifetime long enough to be consistent with
experiment, and lowering the gap between $M_{GUT}$ and the Planck mass $M_{Pl}$.

For these  and other reasons, since the early 1980's, SUSY-GUT's
have been the attractive form of GUT's.  One  reason that I have
not yet mentioned, which is very important even though it only
involves a partial success, has to do with the hierarchy problem.
SUSY stabilizes the hierarchy $M_W/M_X<10^{-13}$, canceling large
radiative corrections to the Higgs boson mass.   A Higgs boson
light compared to the GUT scale is thus made technically natural,
but is not yet explained, and one is also left without an
understanding of doublet-triplet splitting (the fact that the
color triplet partners of the ordinary Higgs doublets are so
heavy compared to the doublets; otherwise there would be very
rapid proton decay).  Also, it was shown (by Alvarez-Gaum\'e,
Polchinski, and Wise) that radiative symmetry breaking would lead
to a natural mechanism of electroweak symmetry breaking if the
top quark is heavy enough. The requisite top quark mass, however,
seemed bizarre at the time.

To continue the story, we must consider developments involving
extra dimensions and gravity.  It was soon realized that although
one cannot unify three generations in four dimensions, one can
readily do so if one starts above four dimensions.  For example,
I constructed an $SO(12)$ model in six dimensions and an $SO(16)$
model in ten dimensions, in each case getting, after
compactification to four dimensions, three generations of quarks
and leptons from a single irreducible representation of the
unified group in higher dimensions.

A much bigger change came in 1984; with Green-Schwarz anomaly
cancellation and the construction (by Gross, Harvey, Martinec,
and Rohm) of the heterotic string, it became feasible to combine
GUT's with string theory and thus to unify all the forces,
including gravity. In this framework (and assuming at least a
small range of energies in which field theory ideas can be
applied), one has to unify all the observed forces plus three
generations of quarks and leptons plus Higgs bosons in one
SUSY-multiplet -- because that is all there is.

One also has to start in ten dimensions with $E_8\times E_8$ or
$SO(32)$ because those are the only ten-dimensional gauge groups
that are allowed by the  anomaly cancellation mechanism. Of these
two choices, the one that works
 is $E_8\times E_8$.
(This assertion again assumes at least a small range of validity of field theory and was shown
in early work on compactification by Candelas, Horowitz, Strominger, and me.)

So in incorporating gravity and string theory into the picture,
one is forced to continue the GUT chain of eqn. \bonon\ to the
end, and to unify the three generations of quarks and leptons.  We
recall that these two steps did not work in four-dimensional
GUT's.  The model is constructed by starting with $\R^4\times K$,
where $\R^4$ is four-dimensional Minkowski space and, to preserve
four-dimensional supersymmetry, $K$ is a compact six-manifold of
a special sort, a ``Calabi-Yau manifold.''  Then, to obey the
equations of motion, one is forced to introduce vacuum
expectation values or VEV's for gauge fields on $K$, breaking
$E_8$ to a subgroup.  This step works without any contrivance and
is forced on us by the equations of motion.

By making a very simple choice (taking the gauge fields VEV's to
lie in an $SU(3)$, $SU(4)$, or $SU(5)$ subgroup of $E_8$), one
can ensure that the unbroken subgroup of $E_8$ is one of the
usual GUT subgroups $E_6$, $SO(10)$, or $SU(5)$.  And the massless
fermions occur in just the right representations, with the number
of generations being a topological invariant of $K$ (together with
its associated gauge bundle).  But we do not know how to determine
just what the gauge field VEV's should be, and because many
choices are possible for $K$, we also cannot predict the number
of fermion generations.

These models are not really four-dimensional GUT's, since
unification really only occurs in ten dimensions.  The
predictions for fermion quantum numbers, $\sin^2\theta_W$, and
proton decay are similar to those of four-dimensional GUT's.
There are also some differences. I will not survey the
differences systematically, but I will mention a few of them. A
higher-dimensional mechanism to solve the doublet-triplet
splitting problem was pointed out in the original Calabi-Yau
paper and used in many subsequent constructions. (This mechanism
has been reconsidered in the last few years in the
phenomenological literature and will be discussed in tomorrow's
lectures.) Also, in string theory models in which unification is
achieved only in higher dimensions, the usual quantization of
electric charge is generally not obeyed.  There are superheavy
unconfined particles with fractional electric charge, coming from
strings wrapping around uncontractible loops in $K$.  Perhaps
they could serve as a dark matter candidate -- though modern
experimental bounds, from MACRO and elsewhere, put a severe limit
on this possibility. Conversely, the quantum of magnetic charge
is in these models generally larger than the Dirac quantum.

The other thing that happened in passing to string theory was
that the gap between the GUT scale and the Planck scale was
lowered again.  We recall that in the original
Georgi-Quinn-Weinberg computation, the GUT scale was below the
Planck scale by a factor of about $10^4$.  If we place things on
a log scale, as we really should, then this is roughly a 20\%
discrepancy, and including supersymmetry reduced the discrepancy
by about a third.  Incorporating string theory, which for weakly
coupled strings means roughly that  the string scale enters the
formula rather than the Planck scale, turned out to cut roughly
another third off of the original discrepancy.  So, by the
mid-1980's, the logarithmic discrepancy between the GUT and
fundamental scales was cut to about 6\%.  What I think is
important here is that the two improvements in the original
computation -- by including supersymmetry and string theory --
both had independent justifications. They were not introduced
just for this reason. The remaining discrepancy of 6\% is small
but real.  There are interesting ideas about reducing it further
or eliminating it (for example using the strongly coupled
heterotic string instead of the weakly coupled one), but I think
they are not as well motivated as were the enrichment of GUT's by
supersymmetry and string theory.

What has happened since?  One important development is certainly
that oscillations in solar and atmospheric neutrinos have pointed
to neutrino masses (or at least mass differences) in roughly the
range that had been estimated from GUT's twenty years earlier.
Meanwhile, astronomy has given other clues.  The acceleration of
cosmic expansion -- whether ultimately interpreted in terms of a
cosmological constant or a more elaborate mechanism perhaps
involving a light scalar field -- should eventually give an
important clue about SUSY GUT's.  The small anisotropies that
have been observed in the cosmic microwave radiation have for
their simplest interpretation an early inflationary period of the
universe at a scale near the SUSY GUT scale.  So together with
the neutrinos, this may represent two observations of phenomena
involving an energy scale close to that of GUT's, although
alternative interpretations are possible in each case.

Another development is that the top quark mass turned out to be quite large -- as assumed in
supersymmetric models of electroweak symmetry breaking that were formulated in the early 1980's.
And, of course, increasingly precise tests of the standard model have been quite consistent
with the SUSY-based approach to the hierarchy problem, while generally adding to the
challenges faced by other approaches to that problem.

So in short, the GUT-based approach to physics has been attractive since it was first put
forward close to thirty years ago; it has been enriched by new ideas, notably supersymmetry
and strings; and there are real hints that it is on the right track, notably from
$\sin^2\theta_W$ and neutrino masses.

If this approach is right, what may we find at accelerators?  The
Higgs boson really should be in reach, since grand unification
does not work without an elementary Higgs boson, which moreover
should not be too heavy or the standard model will break down at
energies far below the GUT scale (spoiling the SUSY GUT
prediction of $\sin^2\,\theta_W$). Just on this grounds, the Higgs
boson probably should not weigh more than about 200 GeV, as is
suggested in any case by the precision electroweak fits.  If we
take into account supersymmetry, which is also needed for grand
unification, and assume the minimal supersymmetric spectrum, then
the Higgs mass should be below about 130 GeV, and it is tempting
to hope that it may lie at the 115 GeV value hinted at by LEP. If
below 130 GeV, the Higgs should appear at Fermilab if the
projected luminosity is reached, and in any case it should be
seen at the LHC.

We can hope for much more beyond the Higgs. Supersymmetric
particles should be in reach of the LHC -- and maybe of Fermilab
-- since the supersymmetric approach to the hierarchy problem
does not make sense if they are too heavy.

But what would the superworld really look like?  Personally, I do
not think that we have a convincing picture of this, in any
detail, partly because of the problems in conservation of flavor,
CP, and even baryon number. So even though we have a hunch that
supersymmetry will be found, the details, if it is indeed found,
will be surprising, at least to me.

The mystery about what the superworld might actually look like is
part of what makes the search so exciting.  Moreover, exploration
of the superworld will be a long project, because there will be
so many new particles and new interactions to unravel.    It will
require high precision, which hopefully will come from electron
colliders such as TESLA or the LC, as well as requiring the
extreme energy that can be obtained in proton colliders like
Fermilab and the LHC. And interpreting the clues that will come
from those machines will certainly require the best efforts of
theorists.

\end